# On thresholdless lasing features in high-β nitride nanobeam cavities: a quantum optical study

Stefan T. Jagsch[1], Noelia Vico Triviño[2], Frederik Lohof[3], Gordon Callsen[1,2], Stefan Kalinowski[1], Ian M. Rousseau[2], Roy Barzel[3], Jean-François Carlin[2], Frank Jahnke[3], Raphaël Butté[2], Christopher Gies[3], Axel Hoffmann[1], Nicolas Grandjean[2], Stephan Reitzenstein[1*]

[1]Institute of Solid State Physics, Technische Universität Berlin, D-10623 Berlin, Germany

[2] Institute of Physics, École Polytechnique Fédérale de Lausanne,
CH-1015 Lausanne, Switzerland

[3] Institute for Theoretical Physics, University of Bremen, D-28334 Bremen, Germany

---

* Corresponding author: email: stephan.reitzenstein@physik.tu-berlin.de

**Exploring the limits of spontaneous emission coupling is not only one of the central goals in the development of nanolasers, it is also highly relevant regarding future large-scale photonic integration requiring energy-efficient coherent light sources with a small footprint. These studies are accompanied by a vivid debate on how to prove and interpret lasing in the high-$\beta$ regime. We investigate close-to-ideal spontaneous emission coupling in GaN nanobeam lasers grown on silicon. Due to their high optical quality, such nanobeam cavities allow for efficient funneling of spontaneous emission from the quantum well gain material into the laser mode. By performing a comprehensive optical and quantum-optical characterization, supported by microscopic modeling of the nanolasers, we identify high-$\beta$ lasing at room temperature and show a lasing transition in the absence of a threshold nonlinearity at 156 K. This peculiar characteristic is explained in terms of a temperature and excitation power dependent interplay between 0D and 2D gain contributions.**

The search for the limits of semiconductor lasers has initiated the development of micro- and nanolasers with optimized gain material, capable of confining light to nearly diffraction limited volumes [1-3]. Such lasers feature a very high spontaneous emission coupling factor $\beta$, allowing to approach the limiting case of thresholdless lasing [4-11]. With respect to the realization of high-$\beta$ semiconductor nanolasers, 1D photonic crystal nanobeam cavities are excellent candidates, as their design promises small footprint nanolasers [12] combined with an efficient funneling of spontaneous emission into the lasing mode. Since their proposal in 2008, nanobeam cavities have opened up a fast growing field of research with high potential for energy-efficient silicon integrated nanophotonics [12-15] and low power on-chip optical data communication [16]. Electrical integration has been successfully demonstrated [12] and their simple geometry features a close to diffraction limited mode volume ($V \sim (\lambda/2n)^3$) and theoretical quality factors $Q$ exceeding $10^7$ [17]. This also leads to exciting opportunities in fundamental research, ranging from cavity quantum electrodynamics effects in the single emitter regime [18] to optogenetics [19]. Of specific relevance to achieve high-$\beta$ lasing is their cavity mode non-degeneracy and the large mode separation, which allows $\beta$-factors approaching unity [13], [20-22]. Together with the efficient carrier confinement inherent to III-nitrides, this makes them an ideal candidate for studying the peculiarities of high-$\beta$ nanolasers under realistic device conditions (room temperature and ambient atmosphere).

Herein, we present detailed temperature dependent studies of GaN nanobeam lasers grown on a silicon substrate. The nanolasers are based on a single In$_{0.15}$Ga$_{0.85}$N quantum well (QW) as gain material and allow us to demonstrate high-$\beta$ lasing at room temperature using continuous wave (cw) excitation. Second-order (intensity) autocorrelation measurements evidence the onset of lasing via an excitation power density dependent transition in emission statistics from thermal bunching towards the Poisson limit, associated with the stimulated emission of photons, i.e. coherent light. Our measurements are complemented with a microscopic laser theory to access simultaneously the I-O characteristic, zero time delay photon autocorrelation function $g^{(2)}(0)$, coherence time of the emission, and the carrier population functions. This combination provides access to the underlying lasing physics, in particular to the "ideal" autocorrelation function that is not detector limited. By combining the calculated coherence times with the detector resolution we can simulate the measured autocorrelation function in excellent quantitative agreement with our experimental data. By investigating the power dependence of the photon statistics we are able to observe a lasing transition, even in the absence of a threshold-nonlinearity, at a temperature of 156 K. We explore and explain this peculiarity by specific impacts of the gain-dimensionality on the lasing characteristics in a large range of temperatures from 20 K up to room temperature. Of particular interest is a transition region at about 160 K, where additional gain contributions from localized states in the QW lead to a thresholdless input-output (I-O) characteristic, even for $\beta < 1$. Our results provide a comprehensive analysis of high-$\beta$ lasing and possible pitfalls in its interpretation.

**Results**

**High-$\beta$ lasing** The effort to develop low power consuming, i.e. low threshold, nanoscale lasers usually goes hand in hand with the quest to achieve high-$\beta$ lasing. Interestingly though, high-$\beta$ nanolasers do not exhibit an abrupt, phase transition-like, lasing threshold [6]. Instead, high-$\beta$ lasing entails a gradual change in emission properties, including output intensity, linewidth and the transition from thermal to coherent emission, over a wide range of excitation powers [6-9], [23-26]. High-$\beta$ nanolasers should thus not be approached as conventional lasers, but additionally through statistical properties of the emitted radiation [6-7]. We would like to emphasize that the concept of "thresholdless lasing", associated with $\beta = 1$ and the absence of non-radiative losses [8], does not imply a threshold at zero excitation, a concept developed in an early publication [4]. Instead, the gradual transition towards coherent emission always occurs at finite excitation and is thus visible in excitation dependent second-order autocorrelation measurements [6-7], [9]. In practice, most publications on high-$\beta$ lasers still rely solely on I-O characteristics in combination with rate equation fitting [27], in particular when it comes to the study of nanolasers operating at elevated temperatures [13], [20-22], [28]. Thereby, coherence and statistical properties of the emission, which cannot be captured using rate equation modeling, are neglected [6]. In order to preserve a reliable and practically meaningful definition for a lasing threshold, it was repeatedly proposed to rely on statistical properties of the emitted radiation [6-7], [9], [23-24]. The second-order autocorrelation function $g^{(2)}(\tau) = \langle I(t)I(t-\tau)\rangle/\langle I(t)\rangle^2$, where $\tau$ is the delay between photon counting events in both arms of a Hanbury-Brown and Twiss interferometer (HBT), is expected to show an excitation power dependent transition from thermal emission (ideally $g^{(2)}(0) = 2$) to the Poisson limit with $g^{(2)}(0) = 1$ at the onset of lasing. In experiment, a convolution of the correlation function with the detector response function is measured and the thermal emission statistic can only be resolved if the coherence time approaches the detector resolution. Thus, the associated thermal bunching can typically only be observed in the threshold region, where the coherence time is already long enough [23-25]. Measuring solely $g^{(2)}(0) = 1$ above a potential threshold is certainly not a sufficient proof for lasing, as one might merely not be able to resolve the thermal bunching. The important observation is the excitation power dependent transition from thermal to coherent emission. A particular signature of high-β lasing in the $g^{(2)}(0)$ trace is a transition from thermal to coherent emission over a wide range of excitation powers and a possible deviation from the expected value of 2 in the thermal regime [7], [25].

**Nanolaser design and fabrication** We address high-β lasing at elevated temperatures with nanolasers that are composed of an AlN buffer layer and a 3-nm-thick In$_{0.15}$Ga$_{0.85}$N QW embedded in a GaN matrix, which was grown on a Si (111) substrate using metalorganic vapor phase epitaxy. Based on this heterostructure, freestanding nanobeam cavities were subsequently processed by means of e-beam lithography and dry etching techniques (see Ref. [13] for more details). The nanobeams, as shown in Fig. 1a, comprise a photonic crystal mirror on both sides, surrounding a taper region with decreasing hole size towards the cavity center, providing a gentle mode confinement in order to increase $Q$ [14]. In Fig. 1b the intensity profile of the fundamental mode is plotted, as calculated using a 3D finite difference time domain (3D-FDTD) solver, showing that the mode is well confined to the cavity region with a mode volume $V = 0.63(\lambda/n)^3$. Figures 1c-d are top and side-view scanning electron microscope (SEM) images of a typical nanobeam cavity.

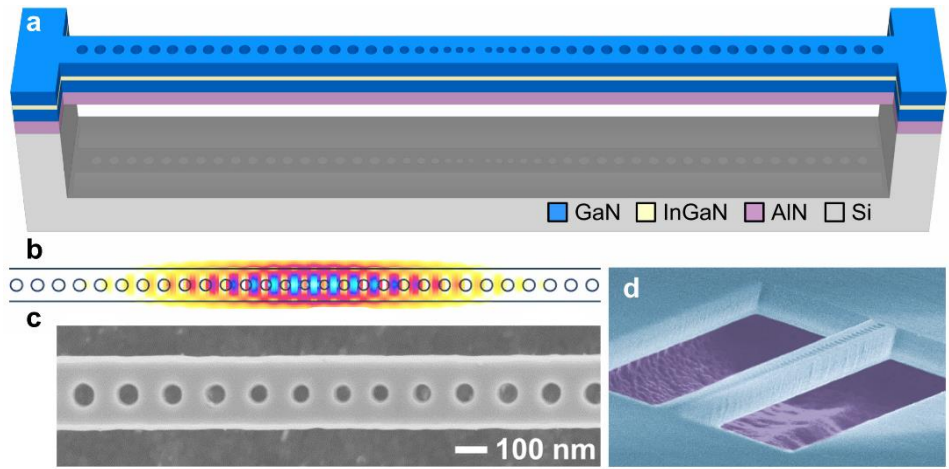

**Figure 1 | III-nitride nanobeam cavity membrane. a**, Schematic drawing of a free-standing nanobeam structure featuring a single InGaN/GaN QW. It consists of two photonic crystal mirrors tapered down to the cavity (not drawn to scale for clarity). **b**, Field intensity profile $|E_y|^2$ of the fundamental cavity mode as obtained via 3D-FDTD simulations. **c**, **d**, Top and side view SEM images of typical nanobeam structures, where the III-nitride layer, the airgap and the silicon underneath (false-color in **d**) are noticeable.

**Room temperature characterization** Optical and quantum-optical characterization of the nanobeam cavities were performed under cw excitation at $\lambda = 325$ nm using a 1/7 chopper wheel to reduce the thermal load. The following evaluation is accompanied by the results of a microscopic laser theory for the interaction between the two-dimensional QW gain material and the fundamental cavity mode. Coupled equations for quantum-mechanical expectation values are solved, describing the wave vector dependent electron and hole populations, the intracavity photon number, as well as correlation functions that connect carrier and photonic degrees of freedom. Including these correlation functions provides direct access to the zero time delay photon autocorrelation function $g^{(2)}(0)$, a quantity that is inaccessible in a rate equation-based analysis. Calculating the first-order coherence function $g^{(1)}(\tau)$ - and with it the coherence time - allows us to model the decay of $g^{(2)}(\tau)$ from its zero time delay value in order to simulate the detector-limited time resolution of the experiment. We would like to emphasize that excellent simultaneous agreement with all experimental data (I-O and $g^{(2)}(0)$ of laser and reference structure) is achieved based on a single set of parameters that enter the model (cf. Table S1 in the supplementary material (SM)). Further details of the optical characterization are provided in the methods summary and the theoretical model is described in detail in the SM.

In Fig. 2 we compare room temperature characteristics representative of a lasing and a non-lasing nanobeam cavity, which serves as a reference. The main difference between both structures lies in the $Q$-factor, which has been determined to $Q{\sim}2200$ for the nanobeam laser and $Q{\sim}1800$ for the reference nanobeam. The room temperature I-O curve of the nanobeam laser is depicted in Fig. 2a, together with the cavity mode below threshold (inset). The solid line is obtained from the microscopic model, assuming a non-radiative loss rate $A_{nr} = 5 \times 10^7 \text{s}^{-1}$, and exhibits a slight threshold nonlinearity before converging to a slope of 1 (dashed line). Due to the strong guiding of photons into the lasing mode, inherent to the nanobeam geometry, emission into non-lasing modes is largely suppressed. This is reflected in a $\beta$-factor of $\sim 0.7$ for both nanobeams, which can be estimated within the scope of the microscopic model along the lines of [10], taking into account the light-matter coupling strength, as well as radiative losses (cf. Table S1). When compared to a rate equation analysis, the ratio of spontaneous emission into the lasing mode is calculated directly, meaning that the $\beta$-factor is no longer an input-parameter (fit parameter) to the theory (see corresponding SM section). Nonetheless, the obtained value is in good agreement with the results of a rate equation analysis of the nanobeam laser ($\beta_{RE} = 0.7 \pm 0.2$) using model and parameters employed in [13]. We conclude that the nonlinear behavior in the I-O curve (Fig. 2a) is mainly caused by non-radiative losses [8]. In comparison, the reference nanobeam (right hand side of Fig. 2) has an I-O curve with a slightly steeper slope of $\sim 1.5$ before converging to 1 and finally saturating (Fig. 2d) before reaching the lasing threshold. The absence of lasing is attributed to slightly higher cavity losses (lower $Q$-factor) and confirmed by autocorrelation measurements and simulation. The I-O curve for the reference nanobeam is well reproduced by theory, accounting for the lower $Q$-factor of 1800, when compared to the lasing case with $Q$ = 2200. Due to the waveguide nature of the cavity, the mode emission in the vertical direction is monitored via scattered light, unless the far-field emission is further optimized using a sidewall Bragg cross-grating outcoupler [29].

It is important to note that soft nonlinearities in the I-O curve, similar to that observed in Fig. 2a, can in principle also be related to trap filling [30] and thus cannot give an unambiguous proof of lasing. In fact, the observation of further indications of stimulated emission is required to prove lasing, in particular towards the limit of a "thresholdless" laser with a linear I-O curve. Another conventional signature of the onset of lasing is a decrease in the emission linewidth at half maximum (FWHM) and an associated increase in temporal coherence at the transition from predominantly spontaneous emission to stimulated emission. In this context, a power dependent linewidth narrowing can also be caused by quenching of absorption losses, which complicates the correct interpretation of this lasing indicator in high-$\beta$ lasers. In contrast to conventional lasers like e.g. vertical cavity surface emitting lasers, where $\beta \ll 1$, high-$\beta$ nanolasers typically show only minor changes in linewidth at threshold [2-3], [23], [27]. This is a result of increased refractive index fluctuations (gain-refractive index coupling) in the excitation range around threshold [26]. Through the soft onset of lasing, which takes place already at low carrier densities and with few intracavity photons [6], these fluctuations persist over a large range of excitation powers, leading to an almost constant linewidth in the threshold region [26]. Considering the discussion above, a lasing threshold could be falsely identified from these classical indicators.

In the present case, we observe a minor decrease in the emission linewidth, superimposed on a dominantly thermal behavior, at $P \approx 5$ kW/cm$^2$, cf. Fig. 2b. Linewidth and emission wavelength of the reference cavity (cf. Fig. 2e) show an overall similar, thermally dominated, trend. A decreasing linewidth at intermediate excitation power densities is not observed. Heating of the freestanding nanobeam membrane occurs at higher excitation powers, the resulting thermal expansion of the nanobeam leading to a redshift of the resonance wavelength with increasing excitation power density (see also Figs. 2b, e). The respective impact on the linewidth is twofold. On the one hand, the expansion of the nanobeam membrane along the temperature gradient from the cavity region reduces the $Q$ factor, which in turn causes an excitation dependent broadening of the cavity mode. On the other hand, fluctuations in the excitation power lead to shifts in the cavity mode position on timescales faster than the integration time, resulting in an additional broadening in photoluminescence. Similar observations were made in Refs. [20] and [22]. The thermal properties of the nanobeams could be improved, however at the cost of reducing $Q$ and $\beta$, by coupling the cavity region directly to the substrate, using e.g. the design adopted in Ref. [12], where a nanopillar under the cavity is used for electrical injection. As the conventionally sought-after lasing signatures (a pronounced I-O nonlinearity and linewidth decrease) are far more elusive, unambiguous proof of the onset of stimulated emission in high-$\beta$ nanolasers requires excitation power dependent second-order autocorrelation measurements in order to monitor the change in emission statistics.

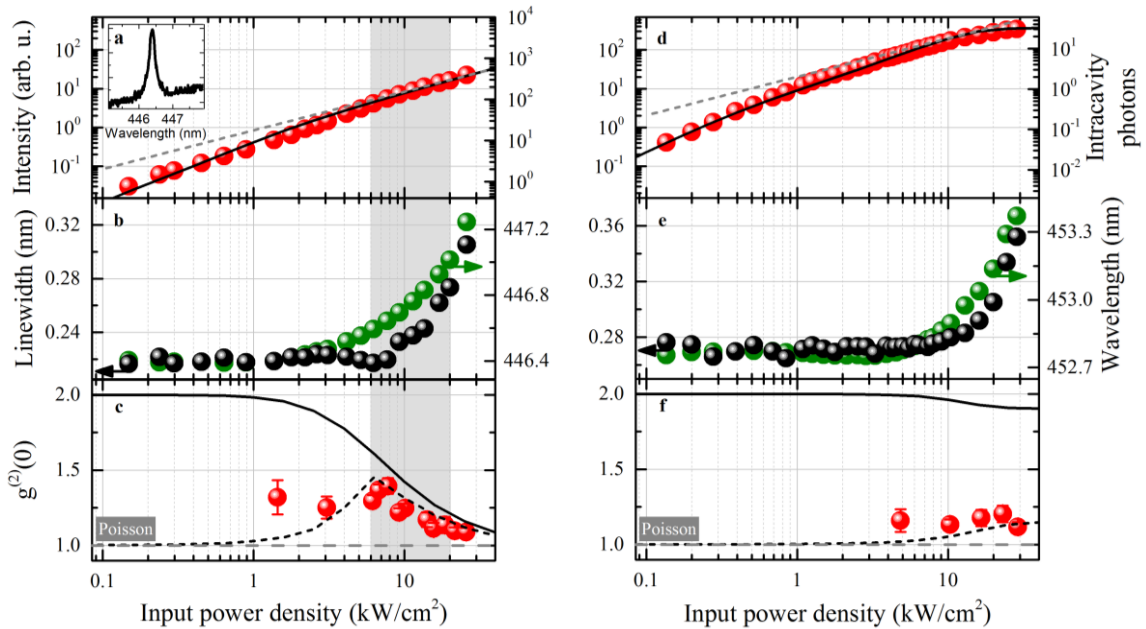

**Figure 2 | Room temperature optical and quantum-optical characterization of a lasing (left) and a non-lasing (right) III-nitride nanobeam cavity. a, d,** Room temperature I-O curves. The theoretical model (solid line) in **a** shows a slight nonlinearity before converging to a slope of 1 (indicated by the dashed line). Inset in **a**: Fundamental cavity mode at 0.64 kW/cm$^2$. The I-O characteristic in **d** is governed by non-radiative losses and

does not show an s-bend before saturating. The increased output intensity with respect to **a** is indicative of an increase in light scattering towards the vertical direction. Note that the intracavity photon number is higher for the lasing nanobeam, which allows building up a coherent photon population. **b**, **e**, Resonance peak wavelength (green) and linewidth (FWHM, black). Above about 10 kW/cm² the development of resonance wavelength and emission linewidth is dominated by heating of the cavity region. The lasing structure in **b** exhibits a slight decrease in linewidth around $P \approx 5$ kW/cm². **c**, **f**, Second-order autocorrelation function at zero time delay as obtained from experiment (data points) and theory. Proof of the transition to coherent emission (shaded excitation range) is provided by the power dependence of the deconvolved second-order autocorrelation data, showing a clear trend towards the Poisson limit ($g^{(2)}(0) = 1$) with increasing excitation power density. In contrast, the power dependence in **f** reveals a constant $g^{(2)}(0) \leq 1.2$. The evolution of the photon statistics is well reproduced by the microscopic theory (ideal: solid line, convoluted: dashed line), when taking into account the calculated coherence time and the convolution with the temporal resolution (~225 ps) of the HBT setup.

The results of an excitation power dependent investigation of the photon statistics are shown in Fig. 2c, f. In order to obtain the zero time delay value $g^{(2)}(0)$, the measured autocorrelation traces were fitted using a convolution of the idealized fitting function and the detector response, taking into account the temporal resolution $\Delta t_{res} \approx 225$ ps of the HBT setup. See also Fig. S3 and corresponding explanations in the SM for further details on the fitting procedure. For the lasing nanobeam, we observe clear bunching behavior ($g^{(2)}(0) > 1$), which becomes less pronounced with increasing excitation power density, indicating the transition from spontaneous to dominantly stimulated emission of light (Fig. 2c). The deduction of a high $\beta$ value is supported by a smeared out and incomplete transition to the Poisson limit within the investigated excitation power density range [25]. The deconvolved data show a bunching maximum of $g^{(2)}(0) \approx 1.4$ in the threshold region. In contrast, in case of the non-lasing cavity, $g^{(2)}(0)$ remains constant ($g^{(2)}(0) \sim 1.2$) over the investigated range of excitation power densities (Fig. 2f). A thermal influence on the photon statistics via a reduced coherence time is not observed. For the nanobeam laser, theory predicts a clear transition from thermal emission to lasing (solid line in Fig. 2c). The transition is accompanied by an increase in coherence time from ~1 ps to ~800 ps across the transition region (cf. Fig. S1 in the SM). Since the decay of the autocorrelation function $g^{(2)}(\tau)$ with respect to $\tau$ is related to the coherence time, the zero time delay value can only be resolved if the coherence time exceeds the detector resolution. This becomes apparent in the convoluted $g^{(2)}(0)$ trace (dashed line in Fig. 2c), which excellently reproduces the experimental results over the excitation range. For the reference nanobeam (Fig.2 f), the experimentally observed constant value of ~1.2 is traced back to an only moderately increasing coherence time, combined with a calculated $g^{(2)}(0)$ that stays largely thermal until saturation is reached. Furthermore, we find spectral hole-burning at the cavity-mode energy in the calculated non-equilibrium carrier distribution functions shown in the SM (cf. Fig. S2a), which is another indicator for lasing operation [10].

Finally, we point out that a mean intracavity photon number of one is not necessarily a signature of lasing [6, 31]. The theoretical results for both the laser and the reference nanobeam exhibit photon numbers above one, but by comparing Fig. 2a and c, one can infer that a coherent photon population builds up only above a mean photon number of 100 [31]. It is the imbalance of spontaneous towards stimulated emission that makes the emission coherent above threshold, and this imbalance is not only determined by the mean intracavity photon number, but also by the inversion of the system that is expressed by the non-equilibrium distribution functions for electrons and holes.

**Temperature dependence** In the following, we address the importance of temperature dependent studies to fully explore and correctly interpret the output characteristics of high-$\beta$ nanolasers. In particular, we analyze and discuss the temperature dependence of carrier confinement and non-radiative recombination and their impact on the I-O characteristics. Upon decreasing the sample temperature, we effectively quench non-radiative loss channels so that the kink in the s-shaped I-O curve becomes less pronounced [8]. Depending on the weight of radiative and non-radiative channels, either a soft s-shape or a more pronounced kink in the I-O curve can be observed and modeled for the same β-factor [8]. As expected, we observe a reduced nonlinearity with decreasing temperature, until approximately 160 K. Below 160 K the I-O curve changes from the familiar s-shape towards an inverse s-shape at low temperatures, exhibiting a completely linear behavior around 156 K (cf. Fig. 3a). This thresholdless I-O curve can be observed despite an extracted β-factor below 1. We model the I-O characteristics at 156 K assuming radiative losses are the same as at RT and non-radiative losses no longer play a significant role ($A_{nr} = 0$). The remaining nonlinearity in the modelled I-O curve originates from the small amount of radiative losses associated with the high, but non-unity, $\beta$ factor.

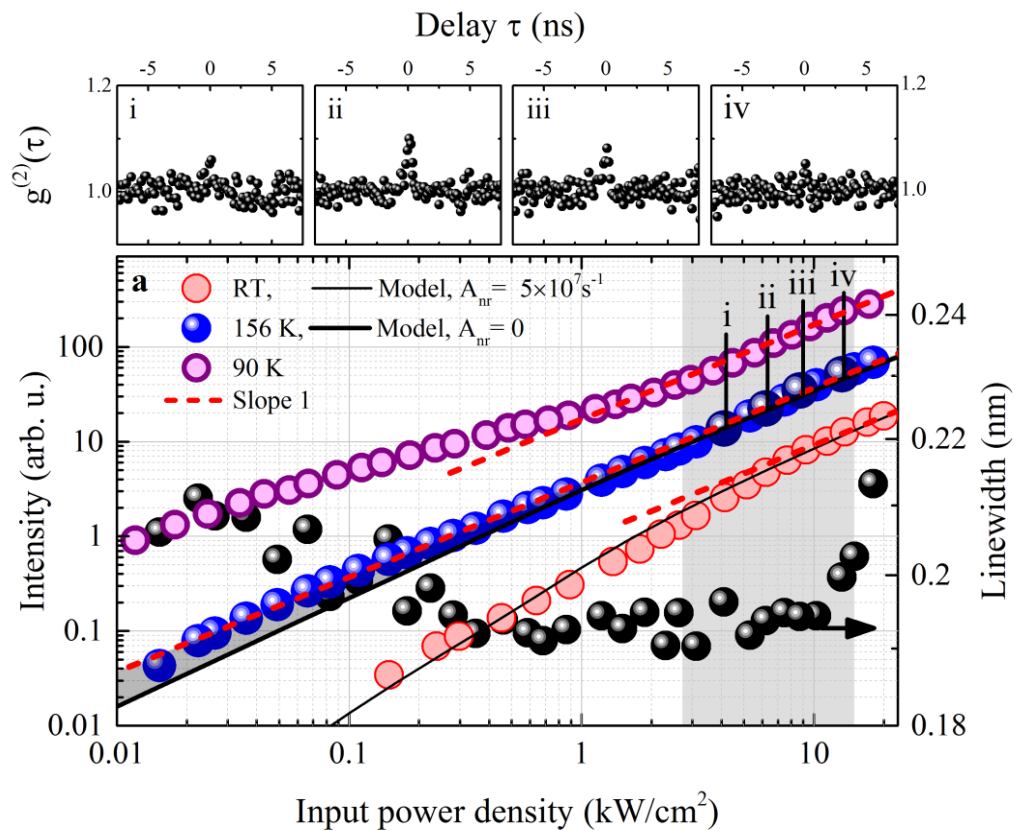

**Figure 3 | Optical and quantum-optical characterization of the thresholdless nanobeam emission at a temperature of 156 K. a**, Excitation power dependent I-O

curve measured at 156 K (blue) exhibiting a thresholdless behavior and corresponding emission linewidth (black dots). I-O curves measured at room temperature (red) and 90 K (purple) show the development of the I-O characteristics with temperature. Results of our microscopic model, based on a purely 2D QW gain, are shown for RT and 156 K data (solid lines). The I-O curve at 156 K is shown under the assumption of negligible non-radiative losses (non-linearity solely due to $\beta < 1$). The increased output intensity in the low excitation regime, due to contributions from localized states, is apparent at 156 K (shaded). Second-order autocorrelation measurements were performed at 156 K at the data points marked **i-iv**. A slope of 1 is indicated by the red dashed lines. The vertical offset was adjusted for clarity. **i-iv**, Autocorrelation traces taken at the excitation power densities marked in **a** display the characteristic bunching around threshold (**ii**, **iii**), which vanishes again for higher excitation (**iv**), indicating a transition towards Poissonian emission statistics in the lasing regime (shaded excitation range in **a**). The bunching in **i** cannot be fully resolved due to resolution limitation.

Instead of ideal spontaneous emission coupling, the thresholdless appearance of the measured I-O curve at 156 K can be ascribed to additional gain contributions from weakly localized states (0D) in the InGaN QW [32-33]. These are activated below a localization temperature of ~160 K, as obtained from an Arrhenius evaluation of the QW emission (see also Figs. S4 and S5 and the related discussion in the SM). The result is a two component 0D-2D gain material. With falling temperature, thermal escape becomes less likely and the number of available localized states increases, leading to increased 0D gain contributions in the low excitation regime (until the localized states are saturated), eventually resulting in an inverse s-shaped I-O characteristic at low temperatures. Around 156 K, these additional gain contributions exactly compensate the threshold nonlinearity and explain the discrepancies between the I-O data and the temperature dependent modeling under the assumption of a purely 2D gain material (cf. Fig 3a).

As for the room temperature case, the emission linewidth associated with the thresholdless I-O curve at 156 K does not show any pronounced narrowing, which could indicate the transition from spontaneous to stimulated emission. Obviously, a threshold can no longer be identified from the sole I-O characteristics. Thus, Fig. 3a is an excellent example demonstrating that a quantum-optical characterization is required to prove possible lasing in high-$\beta$ nanolasers. We performed such measurements for four excitation power densities, as marked in Fig. 3a. Here, clear bunching at zero time delay appears in the autocorrelation trace and vanishes as the excitation power is further increased (cf. Figs. 3i-iv). This unambiguously proves the transition in photon statistics at the onset of lasing and highlights the importance of a quantum-optical investigation to demonstrate lasing in the $\beta \rightarrow 1$ regime. In comparison to the room temperature measurements, we can only observe the far end of the bunching regime due to resolution limitation (cf. Fig. 3i). The visibility of the bunching in $g^{(2)}(\tau)$ is also influenced by the characteristic timescale of intensity fluctuations in the few photon regime (low excitation data points in Fig. 2c, f) [6-7], [25], which changes with temperature and is smaller at 156 K in the present system.

In conclusion, we provide a comprehensive study of high-$\beta$ lasing under realistic device conditions (room temperature and ambient atmosphere) of GaN based nanobeam cavities on silicon, which is substantiated by microscopic laser theory. By combining theory and experiment, lasing is unambiguously identified through the simultaneous observation of a transition to coherent emission in the second-order photon correlation function, an increase of the coherence time, spectral hole burning in the carrier population functions and a slight threshold nonlinearity in the I-O characteristics. Upon decreasing temperature we observe thresholdless lasing at 156 K, which is identified by an excitation dependent quantum-optical characterization of the emission statistics. We highlight the importance of analyzing the temperature dependent carrier confinement and the dimensionality of the gain medium to correctly understand and interpret the characteristics of semiconductor nanolasers. We exemplarily show that the different temperature dependence of 0D and 2D gain media crucially impacts the performance of our nanobeam lasers, mimicking thresholdless lasing at a temperature of 156 K despite a $\beta$-factor below 1. Our results give important insights into the manifold peculiarities of semiconductor nanolasers and highlight central issues and pitfalls in the study of high-$\beta$ lasing. To this end, we show that the photon statistics of emission remain a sensitive indicator of a lasing transition, with particular importance in the high-$\beta$ limit.

## Methods

**Experimental setup.** The 325 nm (3.81 eV) emission line of a HeCd laser was applied for above bandgap cw excitation of the GaN matrix using a 1/7 chopper at 200 Hz in order to reduce the thermal load at high excitation powers. A 20x long working distance UV objective (NA=0.4) was employed for excitation and emission collection in a combined micro-photoluminescence (µ-PL) and Hanbury-Brown and Twiss (HBT) second-order autocorrelation setup. The sample was mounted in a helium flow cryostat. For the µ-PL measurements, the luminescence was dispersed by a single monochromator equipped with a charge-coupled device array. The optical resolution is better than 200 µeV at a photon energy of 2.75 eV. All resulting spectra were calibrated with a mercury gas discharge lamp. For the second-order autocorrelation measurements, the luminescence was guided through a monochromator with an optical resolution of 500 µeV onto a non-polarizing beam splitter and detected by two bialkali photomultiplier tubes in HBT configuration. The measured temporal resolution of the setup is $\Delta t_{res} \approx 225$ ps. Conventional photon counting electronics were used to obtain the final histograms that mirror the second-order autocorrelation function $g^{(2)}(\tau)$ of the nanobeam emission. Room temperature measurements were carried out in a nitrogen atmosphere in order to reduce excitation induced surface depositions over time. All measurements below room temperature were conducted in a controlled low pressure helium atmosphere in order to avoid excessive sample heating present under vacuum.

**Acknowledgements** The authors acknowledge the contributions of Irene Sánchez-Arribas for the optical characterization of the unprocessed sample. The research leading to these results has received funding from the German Research Foundation via the D-A-CH project Re2974/8-1, and projects Re2974/10-1, Gi1121/1-1, the European Research Council under the European Union's Seventh Framework ERC Grant Agreement No. 615613, the Swiss National Science Foundation Grants Nos 200020-113542, 200020-150202 and 200020-162657 as well as the DACH-FNS Grant 200021E-154668.

**Author contributions** N.G., A.H. and S.R. initiated the research and conceived the experiments. S.J., G.C and S.K. performed the experiments. S.J. and S.R. performed the data analysis and the numerical modeling. J.-F.C. grew the samples which were subsequently processed by N.V.T. following extensive 3D-FDTD simulations. I.M.R. and R.B. performed the optical analysis of the unprocessed sample and wrote the corresponding SM sections. F.L., R. Barzel, F.J. and C.G. developed the microscopic model. F.L. and R. Barzel performed the simulations. S.J., C.G. and S.R. wrote the manuscript with contributions from all other authors.

**Author Information** The authors declare no competing financial interests.

**Supplementary Material:**

**On thresholdless lasing features in high-β nitride nanobeam cavities: a quantum optical study**

Stefan T. Jagsch[1], Noelia Vico Triviño[2], Frederik Lohof[3], Gordon Callsen[1,2], Stefan Kalinowski[1], Ian M. Rousseau[2], Roy Barzel[3], Jean-François Carlin[2], Frank Jahnke[3], Raphaël Butté[2], Christopher Gies[3], Axel Hoffmann[1], Nicolas Grandjean[2], Stephan Reitzenstein[1]

[1]Institute of Solid State Physics, Technische Universität Berlin, D-10623 Berlin, Germany

[2] Institute of Physics, École Polytechnique Fédérale de Lausanne, CH-1015 Lausanne, Switzerland

[3] Institute for Theoretical Physics, University of Bremen, D-28334 Bremen, Germany

## A. Microscopic Laser Model

Our theoretical analysis comprises the input-output characteristics, the zero delay time second-order photon correlation function $g^{(2)}(\tau = 0)$ and the coherence time obtained from a two-time calculation of the first-order autocorrelation function $g^{(1)}(t, \tau)$. The model accounts for quasi-continuous $k$-states of the two-dimensional quantum well (QW) gain material, the lowest-energy mode that is provided by the cavity and their mutual interaction. We use a quantized light field in order to access the statistical properties of the emission via $g^{(2)}(\tau = 0)$ and to naturally include spontaneous emission in our model. The operators $b^\dagger$ and $b$ create or annihilate a photon in the laser mode and operators $v_k$ and $c_k$ refer to carriers in the valence- and conduction-band states of the gain material. In this notation the Hamiltonian of the system is written as

$$H = H_\text{carr} + H_\text{ph} + H_\text{I}, \tag{1}$$

where $H_\text{carr}$ and $H_\text{ph}$ are the Hamiltonians of the charge carriers and photons

$$H_{carr} = \sum_k \varepsilon_k^e c_k^\dagger c_k + \sum_k \varepsilon_k^h v_k^\dagger v_k, \tag{2}$$

$$H_{ph} = \hbar\omega \left(b^\dagger b + \frac{1}{2}\right), \tag{3}$$

and $H_I$ is the interaction Hamiltonian

$$H_I = i \sum_k \left(g b v_k c_k^\dagger - g^* b^\dagger v_k^\dagger c_k\right). \tag{4}$$

Energies $\varepsilon_k^\text{e}$ and $\varepsilon_k^\text{h}$ are the electron and hole energies for different momenta $k$ and $\hbar\omega$ is the photon energy of the cavity mode. For a solid quantum well, as it is found e.g. in a VCSEL structure, we define the light-matter coupling constant $g$ as

$$g = \sum_{\mathbf{q}_\parallel} 2E_{\text{ph}}\Gamma_z \mathbf{d}_{cv}\tilde{\mathbf{u}}(\mathbf{q}_\parallel). \tag{5}$$

$E_{\text{ph}} = \sqrt{\frac{\hbar\omega}{2\varepsilon_0 V_{\text{res}}}}$ is the field per photon in the cavity mode with $V_{\text{res}}$ the resonator volume, $\Gamma_z$ the confinement factor in z-direction (vertical to the QW), $\mathbf{d}_{cv}$ the dipole moment between valence and conduction band and $\tilde{\mathbf{u}}(\mathbf{q}_\parallel)$ the Fourier transform of the (cavity-) mode function inside the QW plane.

An approximate relation between the light-matter coupling constant and the spontaneous emission time can be established by adiabatically solving equation (10) and inserting it into equations (8), (9) while neglecting all terms but the one corresponding to spontaneous emission into the lasing mode and only considering the $k$-value in spectral resonance with the optical mode. For $k$-values other than the ones close to resonance, the detuning reduces the light-matter interaction. One arrives at the expression

$$g = \sqrt{\frac{\kappa + \Gamma}{2\tau_l}} \tag{6}$$

where $\kappa$ is the inverse cavity lifetime, $\Gamma$ is the dephasing rate and $\tau_l$ is the spontaneous emission time into the cavity.

In addition, dissipative processes enter the following equations of motion via Lindblad terms.

**Coupled Laser Equations**

Using Heisenberg's equation of motion and a truncation of the arising hierarchy of coupled equations along the lines of Ref. [1], we arrive at dynamical equations for the mean photon number $\langle b^\dagger b \rangle$ and the carrier-distribution functions $f_k^e, f_k^h$ (for electrons e and holes h separately),

$$\left(\hbar \frac{d}{dt} + 2\kappa\right)\langle b^\dagger b \rangle = 2|g|^2 \sum_{k'} \text{Re}[\langle b^\dagger v_{k'}^\dagger c_{k'}\rangle], \tag{7}$$

$$\hbar \frac{d}{dt} f_k^e = -2|g|^2 \text{Re}[\langle b^\dagger v_k^\dagger c_k \rangle] - \gamma_{\text{nl}} f_k^e f_k^h - \gamma_{\text{rel}}^e (f_k^e - f_k^{\text{F.D.}}) - A_{\text{nr}} f_k^e + \frac{P f_k^{(0)}}{n_P}(1 - f_k^e), \tag{8}$$

$$\hbar \frac{d}{dt} f_k^h = -2|g|^2 \text{Re}[\langle b^\dagger v_k^\dagger c_k \rangle] - \gamma_{\text{nl}} f_k^e f_k^h - \gamma_{\text{rel}}^h (f_k^h - f_k^{\text{F.D.}}) - A_{\text{nr}} f_k^h + \frac{P f_k^{(0)}}{n_P}(1 - f_k^h). \tag{9}$$

Their dynamics are governed by the photon-assisted polarization $\langle b^\dagger v_k^\dagger c_k \rangle$ that describes the process of photon emission via a carrier transition from a conduction-band to a valence-band state $k$. Photon losses expressed by the $Q$-factor of the mode are accounted for by the cavity loss rate $\kappa = E_{\text{ph}}/Q$. The carrier dynamics is subject to both radiative and non-radiative losses at rates $\gamma_{\text{nl}}$ and $A_{\text{nr}}$ as well as pumping.

To simulate the experimental situation of carrier excitation in the barrier material the subsequent capture into the QW quasi-continuum states, we assume a carrier generation that is Gaussian distributed higher above the band edge from where carriers relax to the band edges. Carrier relaxation towards quasi-equilibrium is treated in terms of a relaxation-time approximation against respective Fermi-Dirac distributions $f_k^{\text{F.D.}}$ for electrons and holes at rates $\gamma_{\text{rel}}^e$ and $\gamma_{\text{rel}}^h$. The carrier distribution functions $f_k^e, f_k^h$ that enter the theory are generally non-equilibrium distributions and reflect effects such as hole burning at the cavity-mode energy in the presence of stimulated emission, which is indicative for lasing and discussed in the context of Fig. S2 below.

In standard laser theory, radiative losses are typically associated with the β factor, which is a measure for the fraction of the spontaneous emission directed into the laser mode. At the same time, the β factor relates to the occurrence of a jump in the input-output curve. If radiative losses are small in comparison to emission into the laser mode, the β factor can approach unity and the input-output curve becomes thresholdless. As radiative recombination requires the presence of an electron and a hole, radiative losses are proportional to $f_k^e f_k^h$. Non-radiative loses, on the other hand, arise from different physical effects, such as Shockley-Read-Hall recombination or Auger processes. We use a general loss rate $A_{\text{nr}}$ that is proportional to the respective carrier population functions $f_k^e$ and $f_k^h$ [2]. As a general property, non-radiative losses lead to an increased slope - larger than one - in the double logarithmic input-output curve in the low-excitation regime. Furthermore, it is worth noting that both, radiative and non-radiative losses, if sufficiently strong, give rise to a jump in the input-output curve and cannot easily be separated without further investigations.

The non-Markovian equation for the photon-assisted polarization is the central quantity that contains the light-matter interaction of the gain material with photons in the laser mode. Its equation of motion is given by

$$\left(\hbar \frac{d}{dt} + \kappa + \Gamma\right) \langle b^\dagger v_k^\dagger c_k \rangle = -i(\varepsilon_k^e - \varepsilon_k^h - \hbar\omega) \langle b^\dagger v_k^\dagger c_k \rangle$$
$$+ f_k^e f_k^h + \langle b^\dagger b \rangle (f_k^e + f_k^h - 1) + \delta\langle b^\dagger b c_k^\dagger c_k \rangle - \delta\langle b^\dagger b v_k^\dagger v_k \rangle. \qquad (10)$$

The light-matter interaction Hamiltonian gives rise to spontaneous emission proportional to $f_k^e f_k^h$, and stimulated processes proportional to $\langle b^\dagger b \rangle (f_k^e + f_k^h - 1)$ plus correlation terms due to carrier-photon correlation functions $\delta\langle b^\dagger b c_k^\dagger c_k \rangle$ and $\delta\langle b^\dagger b v_k^\dagger v_k \rangle$. Depending on the sign of the population term, it represents stimulated emission (gain) or absorption proportional to the mean intra-cavity photon number. The free evolution of this equation is governed by the detuning of each electronic transition $\varepsilon_k^e - \varepsilon_k^h$ to the cavity-mode energy $\hbar\omega$. The bandwidth of the interaction is determined by the broadening of the cavity-mode resonance (passive cavity *Q*-factor) and the linewidth of the gain material that we describe by the phenomenological constant $\Gamma$ (representing the dephasing rate of the gain medium).

The carrier-photon correlation terms $\delta\langle b^\dagger b c_k^\dagger c_k\rangle$ and $\delta\langle b^\dagger b v_k^\dagger v_k\rangle$ can have a significant impact on the emission characteristics of nanolasers with strong light-matter interaction. Their calculation is a prerequisite to access the photon autocorrelation function

$$g^{(2)}(\tau = 0) = 2 + \frac{\delta\langle b^\dagger b^\dagger b b\rangle}{\langle b^\dagger b\rangle^2},$$

which requires to include four additional equations:

$$\left(\hbar\frac{d}{dt} + 4\kappa\right)\delta\langle b^\dagger b^\dagger b b\rangle = 4|g|^2 \sum_{k'} \mathrm{Re}\left[\langle b^\dagger b^\dagger b v_{k'}^\dagger c_{k'}\rangle\right], \tag{11}$$

$$\begin{aligned}\left(\hbar\frac{d}{dt} + 3\kappa + \Gamma\right)\delta\langle b^\dagger b^\dagger b v_k^\dagger c_k\rangle &= -i(\varepsilon_k^e - \varepsilon_k^h - \hbar\omega_l)\delta\langle b^\dagger b^\dagger b v_k^\dagger c_k\rangle \\ &\quad -2|g|^2\langle b^\dagger v_k^\dagger c_k\rangle^2 - (1 - f_k^e - f_k^h)\delta\langle b^\dagger b^\dagger b b\rangle \\ &\quad +2f_k^h \delta\langle b^\dagger b c_k^\dagger c_k\rangle - 2f_k^e \delta\langle b^\dagger b v_k^\dagger v_k\rangle \\ &\quad +2\langle b^\dagger b\rangle(\delta\langle b^\dagger b c_k^\dagger c_k\rangle - \delta\langle b^\dagger b v_k^\dagger v_k\rangle) \\ &\quad -2\sum_{k'}\delta\langle b^\dagger b c_{k'}^\dagger v_{k'}^\dagger v_k c_k\rangle + \sum_{k'}\delta\langle b^\dagger b^\dagger v_{k'}^\dagger v_k c_{k'} c_k\rangle,\end{aligned} \tag{12}$$

$$\left(\hbar\frac{d}{dt} + 2\kappa\right)\delta\langle b^\dagger b c_k^\dagger c_k\rangle = -2|g|^2\mathrm{Re}\left[\delta\langle b^\dagger b^\dagger b v_k^\dagger c_k\rangle + \sum_{k'}\delta\langle b^\dagger v_{k'}^\dagger c_k^\dagger c_{k'} c_k\rangle + (\langle b^\dagger b\rangle + f_k^e)\langle b^\dagger v_k^\dagger c_k\rangle\right], \tag{13}$$

$$\left(\hbar\frac{d}{dt} + 2\kappa\right)\delta\langle b^\dagger b v_k^\dagger v_k\rangle = 2|g|^2\mathrm{Re}\left[\delta\langle b^\dagger b^\dagger b v_k^\dagger c_k\rangle - \sum_{k'}\delta\langle b^\dagger c_{k'}^\dagger v_k^\dagger v_k v_{k'}\rangle + (\langle b^\dagger b\rangle + f_k^h)\langle b^\dagger v_k^\dagger c_k\rangle\right]. \tag{14}$$

By truncating higher-order correlation functions (see Ref. [1]), Eqs. (7)-(14) form the closed system of coupled laser equations that are used for obtaining the results presented in the main text. We model the quasi-continuous $k$-states by using a discrete grid of ~350 equally spaced points, which leads to a system of about 1400 coupled equations.

**Discussion of the $\beta$ factor**

The $\beta$ factor plays a fundamental role in rate equation theories and is often considered as a central device characteristic. While in rate equations it directly enters as a parameter, the intricate physical processes that determine the behavior of the introduced laser model do not allow one to treat the impact of radiative carrier losses in terms of a single parameter. Nevertheless, a parameter that contains the same physical meaning, i.e. the ratio of the spontaneous emission into the laser mode to the total spontaneous emission can be calculated from our theory by evaluating

$$\beta = \frac{\sum_k \frac{2|g|^2}{\Gamma+\kappa}L(k)f_k^e f_k^h}{\sum_k \left(\frac{2|g|^2}{\Gamma+\kappa}L(k)+\gamma_{\mathrm{nl}}\right)f_k^e f_k^h}, \tag{15}$$

where $L(k)$ is a Lorentzian lineshape function. The expression is derived by adiabatically solving Eq. (10) for the steady state and inserting it in Eqn. (8) and (9) where non-radiative losses are omitted. In the calculation of the emission into the laser mode, stimulated contributions are suppressed, so that the obtained $\beta$ factor truly characterizes the spontaneous emission behavior. The $\beta$ factor of around 0.7 given in the main text is an upper estimate on the basis of expression (15).

From this procedure it becomes clearer that formally differentiating between radiative and non-radiative losses in defining an efficiency-factor such as $\beta$ is rather artificial. Non-radiative losses deplete the carrier populations in Eqs. (8) and (9) in a similar way to radiative losses $\propto \gamma_{\text{nl}}$, and an alternative definition of the $\beta$ factor can be formulated

$$\tilde{\beta} = \frac{\sum_k \frac{2|g|^2}{\Gamma+\kappa} L(k) f_k^e f_k^h}{\sum_k \left(\frac{2|g|^2}{\Gamma+\kappa} L(k) + \gamma_{\text{nl}} + A_{nr}\right) f_k^e f_k^h},$$

in order to characterize the overall efficiency of carrier recombination into the laser mode. With this discussion we aim at clarifying the role of $\beta$, which is very well defined in rate equation theories such as in Ref. [3], but less so in microscopic theories that do not depend on $\beta$, but allow for a definition of such a factor in one way or another.

**Coherence Time**

The coherence time is defined as

$$\tau_c = \int_{-\infty}^{\infty} \frac{|g^{(1)}(t,\tau)|^2}{|g^{(1)}(t,0)|^2} \, d\tau, \qquad (16)$$

which requires calculating the $\tau$-dynamics of the two-time first-order photon correlation function

$$g^{(1)}(t,\tau) = \frac{\langle b^\dagger(t) b(t+\tau)\rangle}{\langle b^\dagger(t) b(t)\rangle}. \qquad (17)$$

For continuous-wave excitation, the first time argument corresponds to the steady-state time. The $\tau$-dynamics is obtained by formulating equations of motion with derivatives taken with respect to the delay-time $\tau$:

$$\hbar \frac{d}{d\tau} G(\tau) = \sum_k g^* P_k(\tau) - (\kappa + i\hbar\omega) G(\tau), \qquad (18)$$

$$\hbar \frac{d}{d\tau} P_k(\tau) = g(f_k^c - f_k^v) G(\tau) - (\Gamma + i(\varepsilon_k^e - \varepsilon_k^h)) P_k(\tau). \qquad (19)$$

Here, the abbreviations

$$G(\tau) = \langle b^\dagger(t) b(t+\tau)\rangle, \qquad (20)$$

$$P_k(\tau) = \langle b^\dagger(t) v_k^\dagger(t+\tau) c_k(t+\tau)\rangle \qquad (21)$$

are used. The initial conditions that enter the time integration of the equations are obtained from the steady-state values of the single-$t$-time dynamics [4].

**Choice of Parameters**

For the unloaded cavity $Q$-factor in the absence of an absorptive gain material we use $Q = 2200$ for the lasing and $Q = 1800$ for the non-lasing nanobeam cavity based on the measurements close to transparency of the gain medium. In all cases, a spontaneous emission time of $\tau_{sp}$ = 5 ns is assumed [5], which leads to a calculated light-matter interaction constant of $g = 0.041$ ps$^{-1}$ at resonance according to Eqn. (6). The cavity loss rates for the investigated nanobeam cavities are $\kappa = 0.94$ ps$^{-1}$ ($Q = 2200$) and $\kappa = 1.15$ ps$^{-1}$ ($Q = 1800$), respectively, and the phenomenological dephasing is $\Gamma = 20$ ps$^{-1}$. For the quantum-well gain material, we estimate an effective area of 0.31 µm² from the 3D-FDTD simulations.

Radiative losses are small in the nanobeam geometry, which is known to strongly funnel photons into the guided laser mode and suppress emission into non-lasing modes. Defect-induced non-radiative losses lead to a deviation of the unity-slope in the input-output curve in the low excitation regime. At 300 K we assume $A_{nr} = 5 \times 10^7$ s$^{-1}$, and an absence of non-radiative losses at 156 K. These parameter choices are supported by the measurements, as a steeper slope than unity is observed at room temperature (especially in Fig. 2d), whereas the 156 K input-output curve is almost flat with a slope of one. The set of parameters is summarized in Table S1.

**Table S1 | Parameters used in the calculation.**

|  | LASING NANOBEAM (RT) | LASING NANOBEAM (156 K) | REFERENCE NANOBEAM (RT) |
|---|---|---|---|
| $Q$ | 2200 | 2200 | 1800 |
| $\tau_{sp}$ (ns) | 5 | 5 | 5 |
| $\beta$ | 0.7 | 0.7 | 0.7 |
| $A_{nr}$ (1/s) | $5 \times 10^7$ | 0.0 | $5 \times 10^7$ |
| $g$ (1/ps) | 0.041 | 0.041 | 0.041 |

**Additional indicators for lasing in the nanobeam structure**

In Fig. S1 we show the results for the coherence time that we obtain for the nanobeam laser and the non-lasing reference structure. While the low-excitation value is mainly determined by the cavity-Q factors, the increase in coherence time is indicative for the build-up of coherence in the system. While the nanolaser reaches coherence-time values is the ns range which is typical for above-threshold operation, the coherence time of the reference structure saturates at a value $< 100$ ps, indicating that fully coherent emission is not reached.

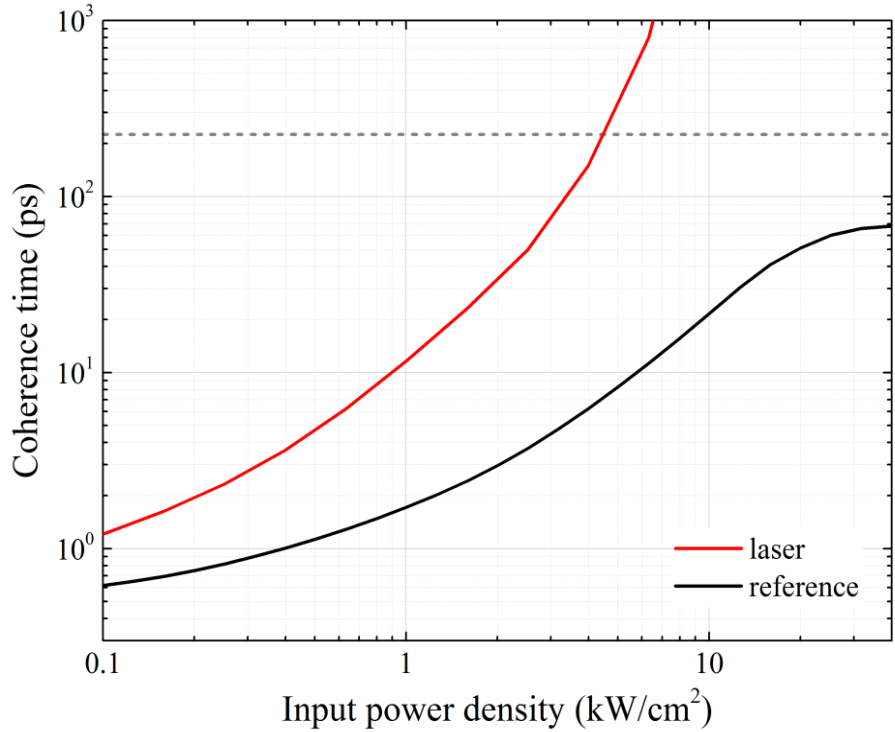

**Figure S1 | Calculated excitation-power-density dependent coherence time.** Excitation dependent coherence time for the nanobeam laser and reference nanobeam according to Eqn. (16). The obtained coherence times are used when convoluting the ideal zero time delay autocorrelation function with the time resolution of the HBT setup (225 ps – shown above with a dashed line) in order to simulate the experimental results.

The coherence time is of particular relevance to predict the measured autocorrelation function (dashed line in Figs. 2c and f in the main text) from on the "ideal" result (solid line in Fig. 2c and f in the main text), which would be obtained only with unlimited detector resolution. The dashed line in Fig. S1 indicates the experimental detector resolution (225 ps). If the coherence is time lower than the detector resolution, the measured autocorrelation function exhibits values that are closer to 1 than the true value.

From Eqns. (7-9) we obtain the wave-vector dependent carrier population functions for electrons and holes. In Fig. S2 we show results for the room temperature case corresponding to the data shown in Fig. 2 in the main text. Spectral hole burning (indicated by the shaded region in Fig. S2 a) gives additional proof for the presence of stimulated emission in the nanolaser in the high-excitation regime and the absence of the same in the reference structure.

In general, the unambiguous identification of lasing in high-$\beta$ nanocavity devices relies on a combination of indicators. The theoretical results for the coherence-time increase and the spectral hole burning in the population functions give valuable additional proof for the presence of lasing in addition to the experimental results that are shown in the main text.

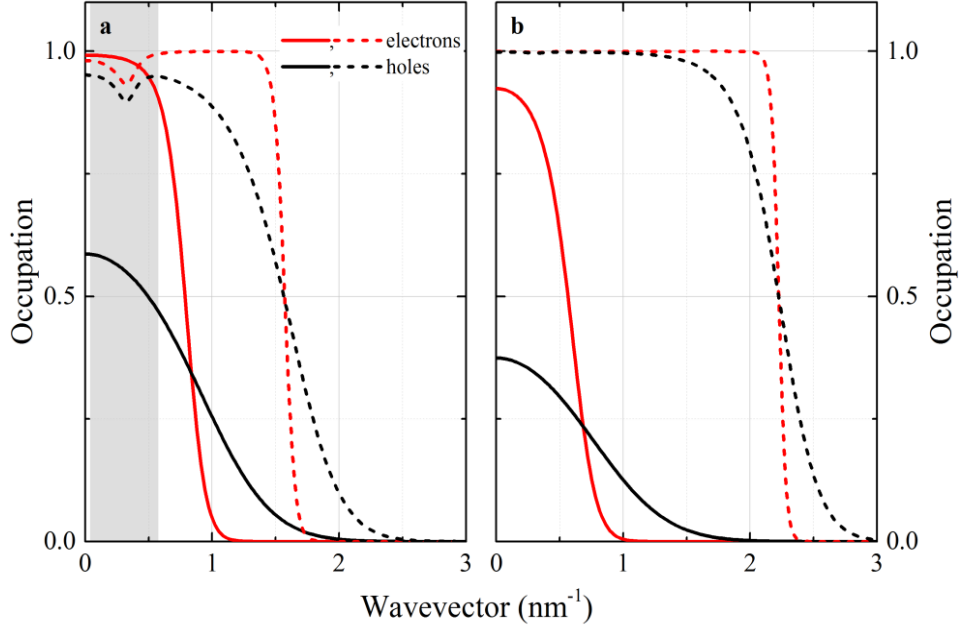

**Figure S2 | Carrier population functions and spectral hole burning.** The non-equilibrium electron (red) and hole (black) distribution functions are shown for low (solid line) and high (dashed line) excitation power density. **a,** The nanobeam laser exhibits spectral hole burning at the laser-mode energy at high excitation (shaded region in **a**), which is indicative for lasing. **b,** No spectral hole burning is observed for the reference structure, confirming that it operates below threshold.

**B. Evaluation of the intensity autocorrelation function**

The autocorrelation traces were fitted using a convolution of the idealized fitting function $g^{(2)}(\tau) = 1 + g_0 \exp(-2|\tau|/\tau_{cor})$, with correlation time $\tau_{cor}$ and bunching amplitude $g_0$, and the detector response (assumed Gaussian) [6]

$$g_{exp}^{(2)}(\tau) = \frac{1}{\sigma\sqrt{2\pi}} \int_{-\infty}^{\infty} g^{(2)}(\tau - \tau') \exp\left(\frac{-\tau'^2}{2\sigma^2}\right) d\tau', \qquad (22)$$

where $\sigma = \Delta t_{res}/2\sqrt{2\ln(2)}$. The temporal resolution of the HBT setup is $\Delta t_{res} \approx 225$ ps. Figure S2a shows the unbinned data (gray) at an excitation power density of 6.77 kW/cm², as well as the data using an 8x binning (black) overlaid with the convoluted fit to the unbinned data (red). Figure S2b shows the evolution of $g^{(2)}(\tau)$ for six excitation power densities around threshold. At $\tau = 0$ we observe a clear bunching behavior ($g^{(2)}(0) > 1$), which becomes less pronounced with increasing excitation power density, indicating the transition from thermal light to coherent light.

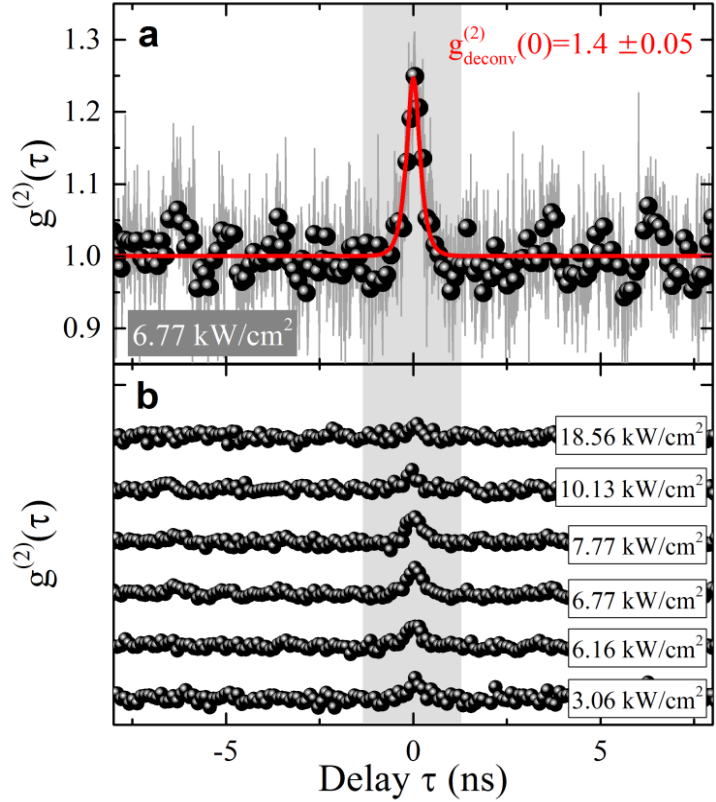

**Figure S3 | Room temperature second-order autocorrelation measurements of the nanobeam emission. a**, Exemplary $g^{(2)}(\tau)$-trace with convoluted fit, Eqn. (22), to the unbinned data (gray) and the data using an 8x binning (black) for clarity. A deconvolved zero delay time value of $g^{(2)}(0) = 1.4 \pm 0.05$ is extracted. **b,** Excitation power density dependent evolution of the second-order autocorrelation function (offset for clarity). As the excitation power density is increased across the threshold region, thermal bunching around $\tau = 0$ (shaded region) becomes more pronounced, before vanishing again at higher excitation power densities as the emission becomes increasingly coherent.

### C. I-O behavior at low temperatures and impact of localized states due to indium composition fluctuations in the single quantum well

We observe a vanishing kink in the I-O curve of the nanobeam laser as the temperature is decreased below room temperature, until the device exhibits thresholdless (linear) I-O behavior despite a cavity β < 1 and the presence of non-radiative losses. Further insight into the lasing characteristics and the mechanism leading to a thresholdless intensity curve is gained via additional temperature dependent measurements below 156 K. Upon decreasing sample temperature, the I-O curve develops an inverse s-shape with an increased output intensity in the low excitation regime (cf. Fig. 3 of the main text), which we observed down to 20 K (not shown). In contrast, a true ($\beta = 1$) thresholdless device requires vanishing non-radiative losses and is expected to remain thresholdless once non-radiative recombinations become negligible. We attribute the I-O temperature dependence to indium composition fluctuations in the InGaN single quantum well (SQW) that lead to a subsystem of localized states [7-8], thereby constituting a 0D-2D two component gain medium. The impact of such localized states has been thoroughly studied for InGaN quantum wells grown on other substrates [7-10]. A typical indicator for the presence of localized states is a blueshift of the QW emission peak energy at intermediate temperatures, following an initial redshift at low temperatures instead of the usual Varshni shift [9-10].

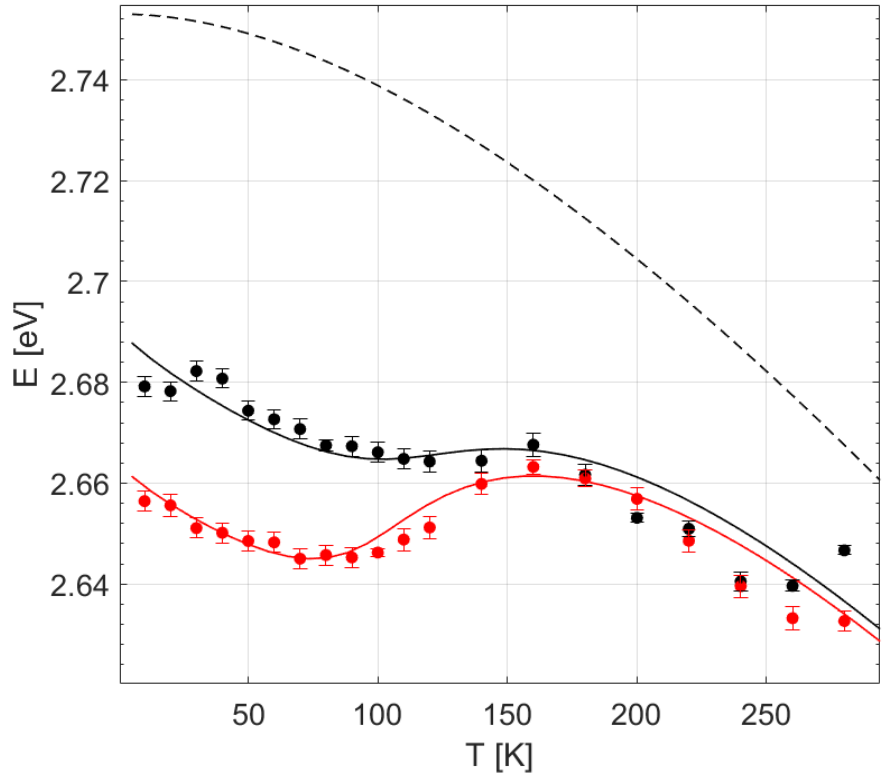

**Figure S4 | Temperature dependence of the InGaN/GaN SQW photoluminescence (PL).** Peak energy extracted from InGaN SQW PL spectra for an unprocessed location on the sample obtained for two low excitation power densities, namely 2.5 (red dots) and 51 (black dots) W/cm$^2$. The blueshift occurring for temperatures above 90-110 K indicates the influence of localized states on the SQW emission. Fits (solid lines) accounting for the temperature dependence of the measured PL peak energy are issued from Eqs. (3) and (4) in Ref. [11] whereas the black dashed line corresponds to the well-known Varshni's empirical formula.

Figure S4 shows the QW emission peak energy for two input power densities as a function of temperature for an unprocessed location on the sample up to $T$ = 280 K, capturing the initial red shift and the subsequent blue shift due to localization starting at 90-110 K. The measurements were performed in standard (macro) PL configuration in the low excitation regime, using the 325 nm line of a HeCd laser at an input power density of 2.5 and 51 W/cm$^2$, respectively. To decrease the inherent light waveguiding originating from the sample geometry, the sample was capped by a $\lambda/4n$ thick (~ 78 nm) SiO$_2$ layer tuned to the QW emission wavelength in order to promote vertical light extraction and hence observe QW PL emission up to 280 K in the low excitation regime. The temperature dependence of the measured PL peak energy can be well accounted for by using the model introduced by Li and co-workers [11] that considers a Gaussian-like distribution of localized electronic states. Within this model the QW emission energy is given by:

$$E_{QW} = E_0 - \frac{\alpha_V T^2}{\beta_V + T} - x(T)k_B T, \tag{23}$$

where $E_0$ is the free QW exciton energy at 0 K, $\alpha_V$ and $\beta_V$ are the usual parameters entering in Varshni's empirical formula, $k_B$ is the Boltzmann constant and $x(T)$ is a dimensionless coefficient obtained when solving the transcendental equation:

$$x \exp x = \left(\frac{\tau_r}{\tau_{tr}}\right)\left[\left(\frac{\sigma'}{k_B T}\right)^2 - x\right]\exp\left[\frac{(E_0 - E_a)}{k_B T}\right], \tag{24}$$

where $\tau_r$ is the carrier recombination time, $\tau_{tr}$ is the carrier transfer time between localized states, $\sigma'$ is the standard deviation of the Gaussian-like distribution of localized electronic states and the energy difference $E_a - E_0$ describes the magnitude of carrier localization at 0 K.

The results of the fitting procedure are summarized hereafter. An energy of $E_0$ = 2.753 eV is deduced for both investigated input power densities. The fact that the measured emission peak energies remain the same for temperatures larger than ~160 K, within the error margin of the measurements, indicates that no screening of the built-in field due to injected carriers occurs. This behavior is fully consistent with the low cw input power densities. Hence the reported emission peak energy blueshift for low temperatures ($T$ < 160 K) between the two input power densities can be ascribed to a progressive filling of localized states. $\sigma$ amounts to 28 ± 0.5 meV for both power densities whereas values of 89 meV and 62 meV are extracted for the energy difference $E_a - E_0$ for the lower and the higher excitation power density, respectively. Additional insights about the bare SQW emission properties can be deduced from the temperature dependence of the integrated PL intensity for the two above-mentioned input power densities (Fig. S4). Thus the larger integrated PL intensity measured for low temperatures for the higher input power density case can be satisfactorily explained in the framework of a faster carrier transfer time combined with a shorter radiative carrier recombination time. The nearly identical value extracted for the thermal activation energy $E_A$ = 80-84 meV for the two investigated input power densities is one more signature that carriers get delocalized above a certain temperature and that they likely experience the same 2D-like potential as also corroborated by their similar emission peak energy.

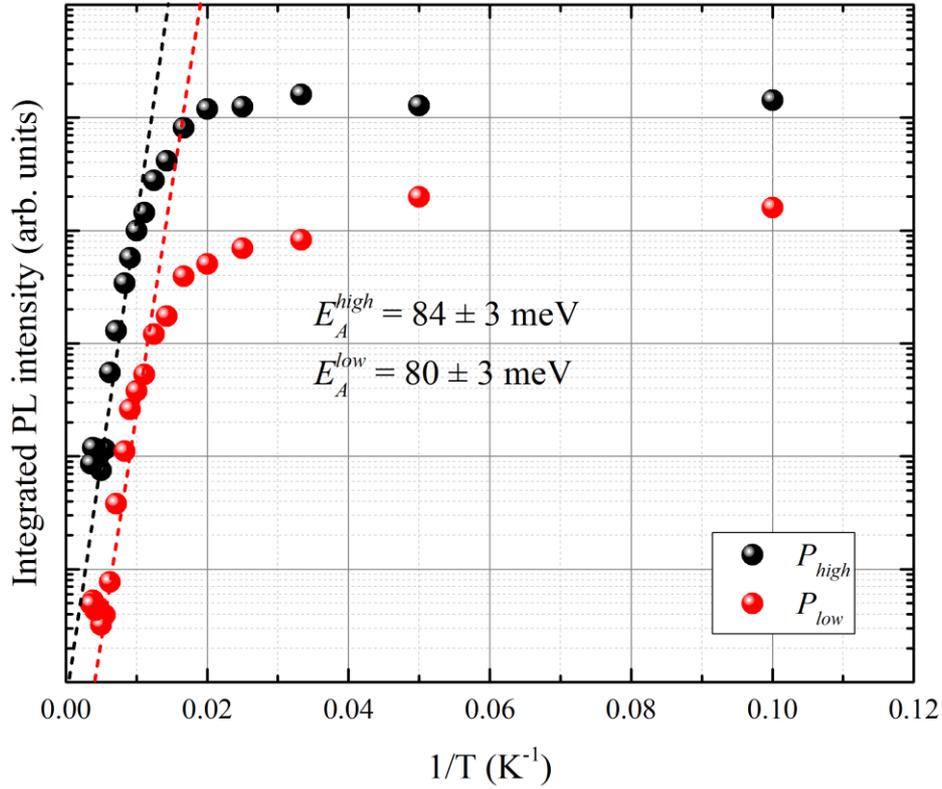

**Figure S5 | Temperature dependence of the integrated PL intensity of the InGaN/GaN SQW.** Temperature dependence of the integrated PL intensity of the bare InGaN/GaN SQW sample obtained for two input power densities, namely 2.5 (red dots) and 51 (black dots) W/cm² together with the corresponding high temperature ($T > 160$ K) thermal activation energy.

When transferring the SQW emission features to the case of the nanobeam laser, at low excitation, 0D-like states are most certainly the main contribution to the gain and the excitation power dependence is linear. As the excitation power density increases, the 0D subsystem saturates and the 2D component increasingly dominates, resulting in an inverse s-shaped I-O curve as a result of the transition from one subsystem to the other. In an intermediate temperature range around 150 K non-radiative losses are exactly compensated by this gain transition, leading to the observed thresholdless I-O curve. We therefore conclude that a thresholdless behavior can be mimicked by a complex gain medium, even in case of a non-ideal spontaneous emission coupling (i.e. $\beta < 1$). Importantly, these effect can lead to an incorrect interpretation of close to linear I-O curves when studying nanolasers only at a single temperature. Against the background provided in part A, this section clearly evidences the need for quantum optical experiments, such as power dependent second-order autocorrelation measurements, as well as a temperature dependent investigation in order to unambiguously assess the transition from spontaneous to stimulated emission and to prove thresholdless behavior in high-$\beta$ nanolasers.